# Electrocatalytic Hydrogen Evolution reaction on edges of a few layer Molybdenum disulfide nanodots

*John Benson[†], Meixian Li[±], Shuangbao Wang[§], Peng Wang[§] and Pagona Papakonstantinou[†]\**

[†]School of Engineering, Engineering Research Institute, Ulster University, Newtownabbey BT37 0QB, UK

[±]College of Chemistry and Molecular Engineering, Peking University, Beijing 100871, P.R.China.

[§]National Laboratory of Solid State Microstructures, College of Engineering and Applied Sciences and Collaborative Innovation Center of Advanced Microstructures, Nanjing University, 22 Hankou Road, Gulou, Nanjing, 210093, P. R. China

ABSTRACT

The design and development of inexpensive highly efficient electrocatalysts for hydrogen production, underpins several emerging clean-energy technologies. In this work, for the first time, molybdenum disulfide ($MoS_2$) nanodots have been synthesized by ionic liquid assisted grinding exfoliation of bulk platelets and isolated by sequential centrifugation. The nanodots have a thickness of up to 7 layers (~4 nm) and an average lateral size smaller than 20 nm. Detailed structural characterization established that the nanodots retained the crystalline quality and low oxidation states of the bulk material. The small lateral size and reduced number of layers provided these nanodots with an easier path for the electron transport and plentiful active sites for the catalysis of hydrogen evolution reaction (HER) in acidic electrolyte. The $MoS_2$ nanodots exhibited good durability and a Tafel slope of 61 mVdec$^{-1}$ with an estimated onset potential of -0.09 V vs RHE, which are considered among the best values achieved for 2H phase. It is envisaged that this work may provide a simplistic route to synthesize a wide range of 2D layered nanodots that have applications in water splitting and other energy related technologies.

**Introduction**

Hydrogen is emerging as an ideal energy carrier for clean and sustainable energy technology[1-4]. Considerable effort has been directed towards the sustainable production of hydrogen from water splitting, which underpins several emerging clean-energy technologies. To be efficient, the process requires the presence of an electrocatalyst, to reduce the overpotential and drive the kinetically rate-limiting steps involved within reductive half reaction of water splitting namely the hydrogen evolution reaction (HER: $2H^+ + 2e^- \rightarrow H_2$). Currently, expensive noble metal platinum is the most efficient electrocatalyst for improving the kinetics and efficiency of HER.

Therefore, the discovery of active HER electrocatalysts made from earth-abundant elements constitutes a key step in the development of large scale hydrogen production technology[5-11].

Molybdenum disulfide ($MoS_2$) compounds have recently attracted considerable attention because of their appealing electrocatalytic properties for electrochemical as well as photochemical hydrogen evolution reaction (HER)[12-18]. These favorable properties emanate from the presence of catalytically active sulfur atoms on the molybdenum edges of $MoS_2$ planes, a high stability in strong acids and an advantageous band gap aligned with the hydrogen redox potential. $MoS_2$ has a layered packed structure consisting of a single layer of Mo atoms sandwiched between two layers of sulfur atoms in a trigonal prismatic arrangement. These stacks are piled in a graphite-like-manner to form bulk material, held together by weak van der Waals forces. Bulk $MoS_2$ is a semiconductor with an indirect bandgap of 1.3 eV, which is modified to a direct band gap semiconductor of 1.8 eV, when it is thinned down to a few layers. Theoretical studies have identified that the unsaturated S atoms located at the edges of $MoS_2$ can absorb H with a small free energy ($\Delta G_H 0 \approx 0.08$ eV)[19,20] and hence act as active sites for HER. This prediction was further confirmed experimentally on $MoS_2$ nanoparticles by systematically varying terrace and edge site densities, where a linear correlation between the exchange current density and $MoS_2$ edge length was established[20]. As a result efforts have been focused on exposing the catalytic edge sites by engineering the morphology of $MoS_2$[21-25]. In this context various strategies have been explored through the development of various nanostructures in the form of ordered architectures, vertically aligned nanosheets, flowerlike nanosheets, microboxes and nanoparticles[26-29].

Although a plethora of active edge sites is beneficial, the vertical charge transport through the layers is an important limitation in HER performance. This needs to be addressed in the design of MoS$_2$ electrocatalysts with optimum performance. Theoretical studies have predicted and experimental studies have confirmed that the electronic structure of MoS$_2$ edges is dominated by metallic one-dimensional edge states[30], which is in sharp contrast to the semiconducting basal plane. As a result the electrocatalysis of atomically thin CVD grown MoS$_2$ films has been found to be strongly dependent on the layer number, which was correlated to the hopping of electrons in the vertical direction of MoS$_2$ layers. These observations led us to the proposition that MoS$_2$ nanodots i.e. a few layer MoS$_2$ sheets with very small lateral dimensions should show enhanced catalytic activity due to a simultaneous increase in unsaturated sulfur edge sites and enhanced electron transfer accrued from the reduced number of layers. Therefore, it is intuitive to construct a scalable method to produce few layer MoS$_2$ to optimize the catalytic performance for HER. To date the HER activity from MoS$_2$ nanodots is largely unexplored, which stems mainly from lack of methods to produce these in appreciable amounts.

Today the synthesis of MoS$_2$ nanodots has been mainly connected with the investigation of photoluminescence arising from quantum confinement effects rather than assessment of HER activity[31-34]. So far top down synthesis of few-layered MoS$_2$ nanodots with lateral dimensions smaller than 50 nm is based either on exfoliation of bulk MoS$_2$ using ultra-sonication in suitable organic solvents or on a combination of grinding and ultrasonication, or on electrochemical exfoliation[24,25,31-34]. Generally, the synthesis of MoS$_2$ nanodots is under intensive investigation and evaluation of their quality and performance for various applications including energy storage, catalysis, gas sensing, and optoelectronic devices, which has been is largely unexplored. Here we report the synthesis of MoS$_2$ nanodots using optimized experimental conditions, that involve grinding MoS$_2$ platelets in a small quantity of room temperature ionic liquid (RTIL) followed by sequential centrifugation steps. As schematically depicted in Figure 1, the combined compressive, torsional and shear forces exerted on the MoS$_2$ platelets are able to exfoliate and at the same time break up the bulk crystals. During the grinding process, the RTIL acts as a lubricant, which allows low friction between the moving pestle and mortar, hence, facilitating the exfoliation. RTILs are salts composed of large, asymmetric and charge delocalized cations that are paired with anions having significantly different chemical structure and symmetry. This incompatibility of the chemical structures inhibits crystallization at room temperature and allows them to have a unique combination of characteristics, including thermal stability, non-flammability and negligible volatility, which make them attractive as lubricants[35].

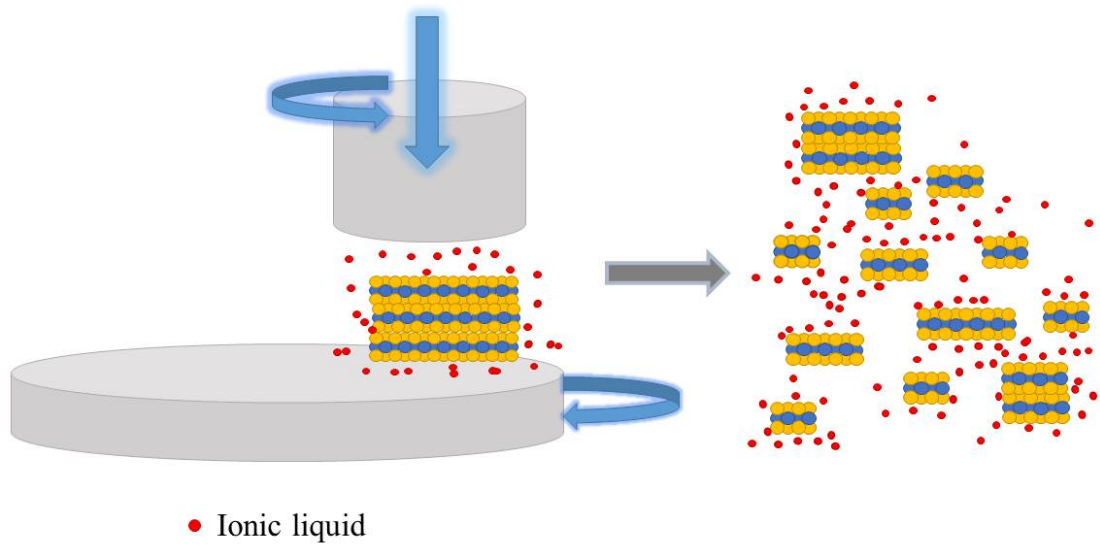

• Ionic liquid

Figure 1. Schematic representation of the ionic liquid assisted grinding exfoliation process. Grinding of MoS$_2$ platelets in the presence of room temperature ionic liquid (RTIL) produces a gel structure, with the ionic liquid acting as a lubricant. Combined compressive, torsional and shear forces exerted on MoS$_2$ platelets in the presence of a small amount of RTIL are able to exfoliate and at the same time break up the bulk crystals.

We show, that the HER activity of nanosized MoS$_2$ is substantially improved, when compared to thicker and larger nanoplatelets. We found that MoS$_2$ nanodots exhibit excellent electrocatalytic activity with low overpotential, mainly attributed to the large number of exposed active conductive edges of $MoS_2$ and to the low number of layers.

**Synthesis**

The synthesis of $MoS_2$ nanodots was carried out in a planetary grinding machine using high purity $MoS_2$ powder with a sufficient amount of RTIL (1-Butyl-3- methylimidazolium hexafluorophosphate, $BMIMPF_6$) to produce a gelatinous material[36-39]. During grinding, the RTIL protects every newly exposed $MoS_2$ surface by adsorbing onto the surface, keeping the sheets separated and avoiding restacking. The adsorption of ionic liquids onto the $MoS_2$ surface is believed to be dominated by weak non-covalent interactions[40].

The resulting gel was subjected to 3 washing steps in a mixture of DMF (N, N-Dimethylformamide) and acetone to remove the RTIL[36]. The formation of $MoS_2$ nanodots is purely a mechanical effect, which arises mainly from the compressive forces exerted during the grinding process. The yield of the nanodots increases by prolonging the duration of grinding. It is worth noting that the mechanical breaking point of $MoS_2$ is 5 times smaller than that of graphene[41] leading to the overall smaller lateral size.

The clean ground product consisted of an assortment of sheets of various sizes and thicknesses, which were size selected by sequential centrifugation as schematically illustrated in Figure 2. Briefly the dispersion of the washed ground product was initially centrifuged at 500 rpm, which resulted in a pellet comprised of large un-exfoliated flakes. The supernatant from 500 rpm run was then centrifuged at 1,000 rpm to isolate sheets with smaller thickness and lateral dimensions. The procedure was repeated using centrifugations at 3,000 and 10,000 rpm, as described earlier[37].

Isolated products are labeled MoS$_2$ XK, where XK represents the centrifugation speed in thousands of rpm.

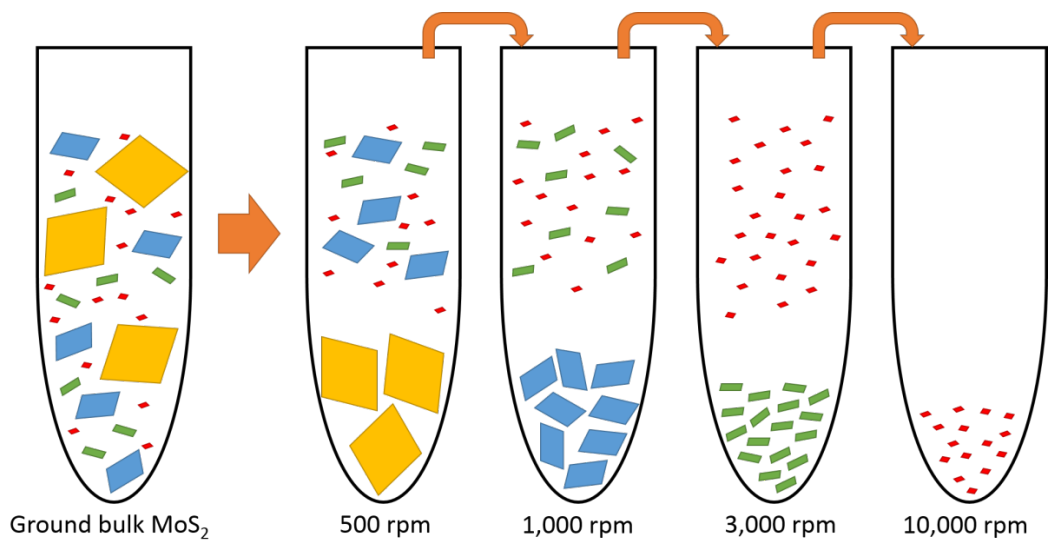

Figure 2. Schematic representation of the sequential centrifugation process. Sequential centrifugation of the supernatant at progressively higher speeds allows the isolation of thinner and smaller particles. Large and thick platelets produce a pellet using low centrifugation speeds and small durations. Thinner and smaller sheets are pelleted at higher speeds and longer durations.

**Results and discussion**

.

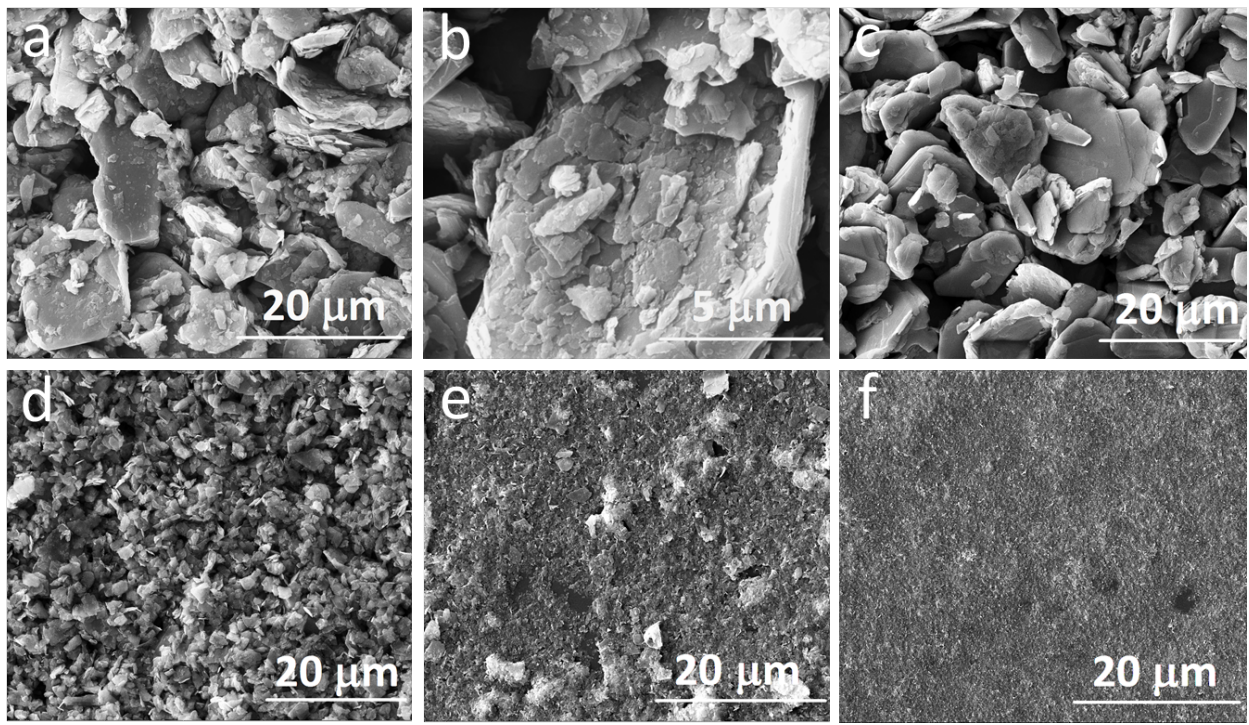

Figure 3. SEM images of isolated sediment: (a) $MoS_2$-bulk, (b) zoomed area of $MoS_2$ bulk (c) $MoS_2$ 0.5K, (d) $MoS_2$ 1K, (e) $MoS_2$ 3K, (f) $MoS_2$ 10K.

SEM images of $MoS_2$ bulk together with those of isolated $MoS_2$ fractions separated at different sequential centrifugation steps are displayed in Figure 3. $MoS_2$ bulk consists predominantly of large platelets (~20μm), as well as a wide range of smaller flake sizes stacked on larger platelets as shown in Figure 3a. Grinding in ionic liquid breaks and exfoliates $MoS_2$ producing a collection of nanosheets of various sizes and thicknesses. Size selection was achieved by subjecting the DMF dispersion of ground washed $MoS_2$ to progressive centrifugation steps, by removing the sediment and subjecting the supernatant to increasing centrifugation speeds. Separation is based on size and density, with larger and thicker (heavier) platelets withdrawing at lower centrifugation speeds (forces) while smaller and lighter particles withdraw at higher speeds. This is evidenced clearly in Figure 3c, where centrifugation at the low speed of 500 rpm withdrew predominantly the large and thick platelets (Figure 3c) leaving in the supernatant an assortment of smaller and lighter material. Subsequent centrifugation of the supernatant at 1000 rpm withdrew ground fragments around 1-6 μm in lateral size, whereas higher centrifugation speeds separated finer material successfully (smaller and thinner) as schematically illustrated in Figure 2. The majority of $MoS_2$ 3K isolated sediment has lateral sizes smaller than 70 nm with a small contribution from larger flakes of about 1-4 μm in lateral size as confirmed by atomic force microscopy (AFM, Figure 4e and Figure 4g). $MoS_2$ 10K isolated sediment is populated with nanodots smaller than 20 nm in size and up to 7 layers in thickness (Figure 4a and Figure S1). The highly parallel and ordered lattice planes in high-resolution TEM (HRTEM) images of $MoS_2$ 10K (Figure 4a) are clearly visible. The d-spacing of 0.3125 nm, which corresponds to the (004) faces of $MoS_2$ crystals indicates that the nanodots retained their crystalline quality after the ionic liquid assisted grinding exfoliation. The clean exfoliated $MoS_2$ nanosheets have a proneness to restack due to the van der Waals interactions. (Figure 4b)[33], which gives rise to multiple sets of rotated spots of variable intensity in the SAED pattern (Figure 4c) and ring motifs. In addition, $MoS_2$ nanodots show a distinct tendency to form agglomerates most probably due to a predisposition to minimize their surface energy by stabilization of the dangling bonds at the edges of the crystals[40]. Overall, the results show that the grinding process allows the high crystalline quality to remain intact in the resulting nanodots. This result is different from our previous work, which reported the presence of amorphous ultra-small $MoS_2$ nanoparticles using ultrasonication in DMF and centrifugation steps[25].

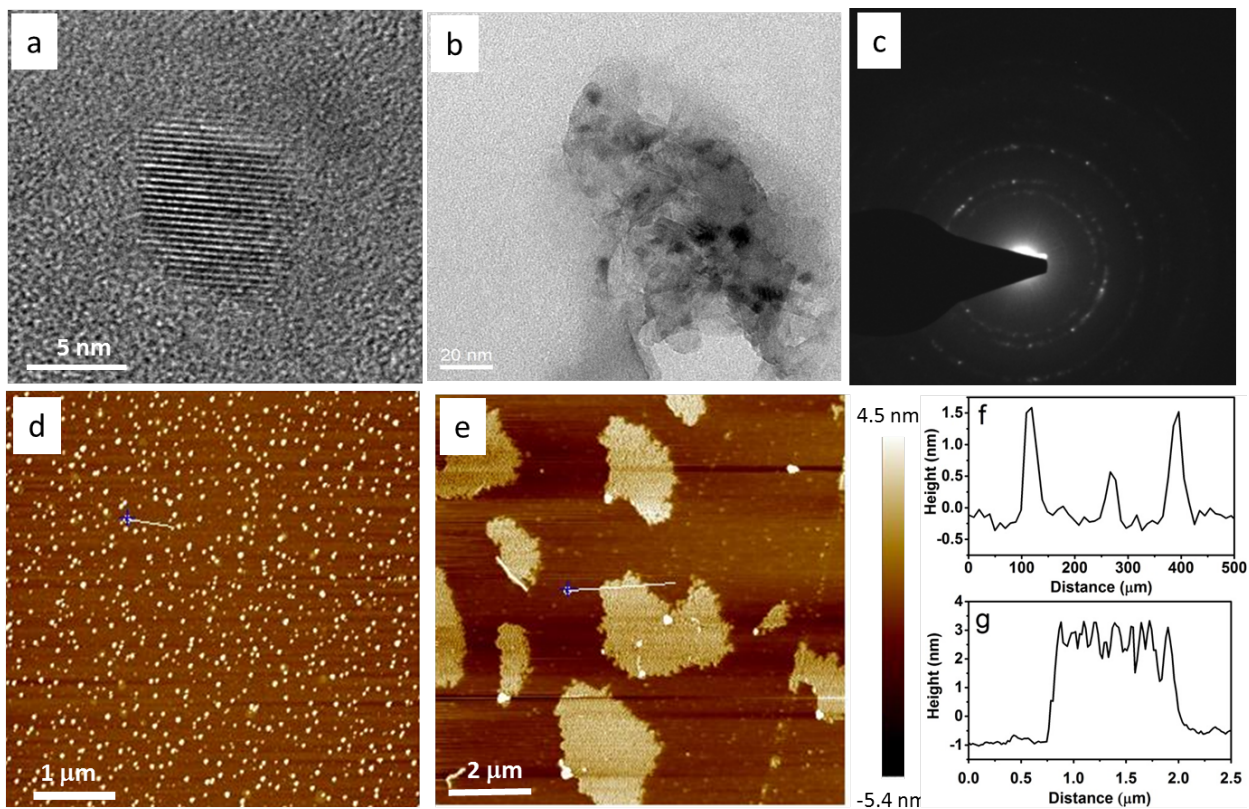

Figure 4. High resolution TEM images of (a) MoS$_2$ 10K nanodot, (b) agglomeration of nanodots, (c) SAED pattern of (b). AFM image of (d) MoS$_2$ 10K and (e) MoS$_2$ 3K with corresponding height profiles shown in (f) and (g), respectively.

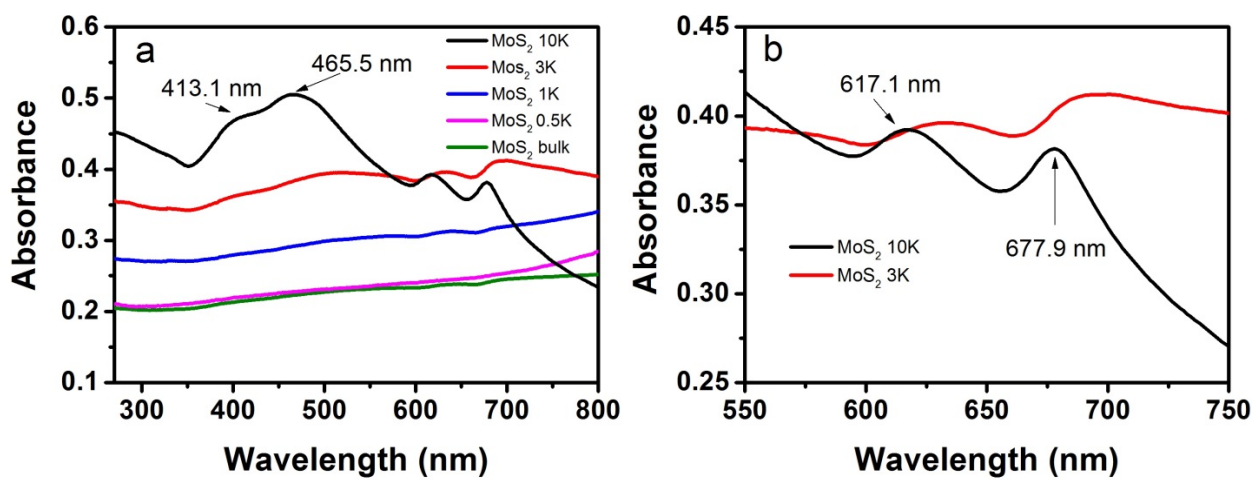

Figure 5. (a) UV-Vis absorption spectrum for all MoS$_2$ products dispersed in DMF. (b) Noticeable blue shift for the MoS$_2$ 10K relative to MoS$_2$ 3K, due to a reduction in thickness and lateral dimensions.

The optical properties of MoS$_2$ centrifugation products dispersed in DMF, were investigated by UV–vis absorption spectroscopy and are shown in Figure 5a. The MoS$_2$ 10K dispersion exhibited four optical absorption peaks at 413.1, 465.5, 617.1, and 677.9 nm, which are characteristic fingerprints of well exfoliated MoS$_2$ nanosheets of 2H type with trigonal prismatic coordination[42,43]. The two well defined peaks centered at 617.1 nm and 677.9 nm, are due to the direct excitonic transitions at the K point of the Brillouin zone[44,45]. The broader peaks at 413.1 and 465.5 nm are assigned to the direct transition from the deep valence band to the conduction band[46]. However, for lower centrifugation speeds the Mie scattering induced background was substantially reduced and the spectra appeared flatten with less distinct peaks; indicating a transition from direct to indirect bandgap, a characteristic of thicker flakes. In addition, MoS$_2$ 10K absorption spectra displayed a blue-shift versus the rest of the products (Figure 5b), which is consistent with quantum confinement effects arising from thickness and lateral size reduction[32].

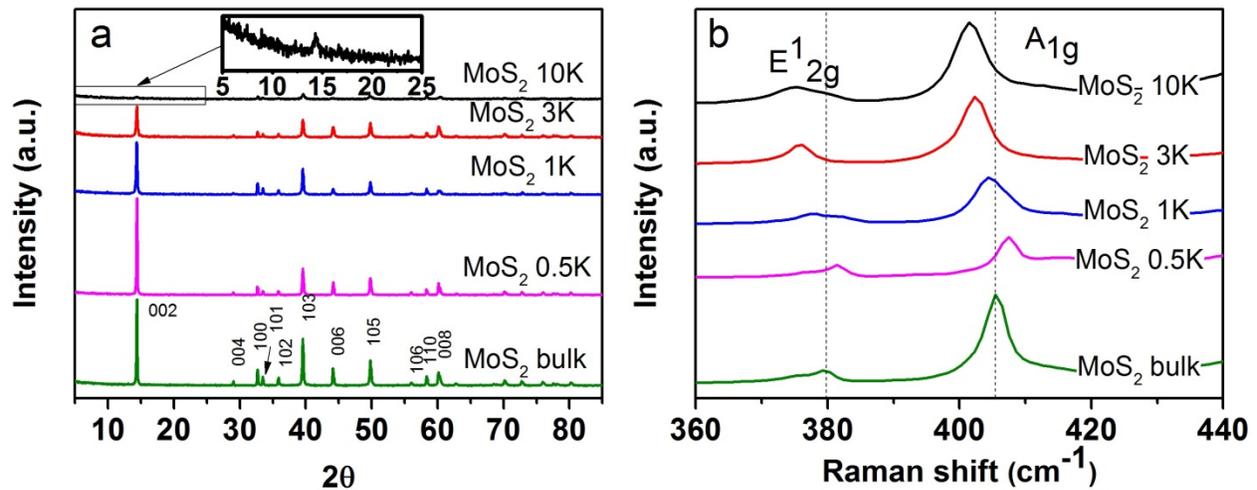

Figure 6. (a) XRD diffraction patterns of $MoS_2$-bulk and isolated sediments. Reduction in (002) peak intensity is associated with progressive exfoliation. (b) Raman spectra were measured using 632.8 nm laser. Significant red shift of $A_{1g}$ mode for increased centrifugation speeds indicates reduction in the number of layers.

X-ray diffraction (XRD) patterns were used to determine the evolution of structural changes at various stages of the sequential centrifugation process (Figure 6a.). $MoS_2$ bulk exhibits a distinctive intense (002) peak, centered at 14.4°, corresponding to an interlayer d-spacing of 0.614 nm. In addition, various weak diffraction reflections from the (100), (103), (006), (105), and (008) planes were observed at higher angles, characteristic of polycrystalline $MoS_2$. No substantial change in the (002) peak was observed for $MoS_2$ 0.5K due to the exposure of large and highly crystalline platelets as revealed by SEM images of Figure 3c. However, higher centrifugation speeds caused a progressive decrease in (002) peak intensity, with $MoS_2$ 10K sediment retaining only 3.4% of the bulk intensity. A similar decrease in the diffraction intensity of the (002) graphite peak was observed in the sequential centrifugation products from ground graphite[37]. Since the diffraction peak intensity is a result of constructive interference from aligned crystal planes, a decrease in the diffraction intensity indicates that a reduction in the number of aligned planes (layers), with zero intensity expected from completely exfoliated $MoS_2$[47,48]. The presence of a weak (002) signal from the $MoS_2$ 10K indicates that the nanodots consist of a few layers, which is in agreement with the AFM results (Figure 4d). In addition a progressive increase in FWHM was observed for centrifugation speeds equal or higher than 1000 rpm (Table S1 supporting information) further confirming the decrease in lateral size.

Raman spectroscopy was used to assess the exfoliation of the products employing a He-Ne laser (632.8 nm excitation wavelength) with a beam size of approximately 2 μm. Figure 6b shows characteristic Raman peaks associated with $E_{2g}$ (in plane motion of Mo and S in opposite directions) and $A_{1g}$ (out of plane motions of S atoms) active modes located at 379.5 and 405.4 cm$^{-1}$, respectively[49-51]. It is well established that the addition of extra layers on $MoS_2$ leads to the stiffening of the out-of-plane phonon modes, resulting in a downward-shift of the A1g mode. Here, the $A_{1g}$ mode is up-shifted by approximately 2.1 cm$^{-1}$ from bulk to $MoS_2$ 0.5K (405.4 cm$^{-1}$ to 407.5 cm$^{-1}$), which indicates that the $MoS_2$ 0.5K platelets are thicker than the bulk ones[51]. Although at first sight this result seems rather counter- intuitive, it can be explained bearing in mind the SEM and XRD results (Figure 3a, Figure 3c and Figure 6a), which show that centrifugation at 500 rpm was very efficient in isolating large and thick platelets, free from small size debris. In contrast, the bulk material besides the large and thick platelets is populated with smaller and finer flakes (Figure 6b), which give rise to the upshift of 2.1 cm$^{-1}$. Centrifugation at speeds higher than 1000 rpm gave rise to a noticeable downshift (404.4 cm$^{-1}$) in the $A_{1g}$ mode indicating a reduction in thickness, as flakes were exfoliated to a few layers. This $A_{1g}$ energy up-shift, suggests that the electron density in exfoliated $MoS_2$ is enhanced most probably due to increased number of edges, which are electron rich[52]. The peak position of $E_{2g}$ mode shifts to smaller wavenumbers and both Raman modes are getting broadened with increasing centrifugation speeds. These variations are in agreement with previous Raman studies on $MoS_2$ quantum dot/nanosheets hybrids[31] and are related to an increase in disorder associated with edge defects since the average nanosheet later size is smaller than the laser spot diameter.

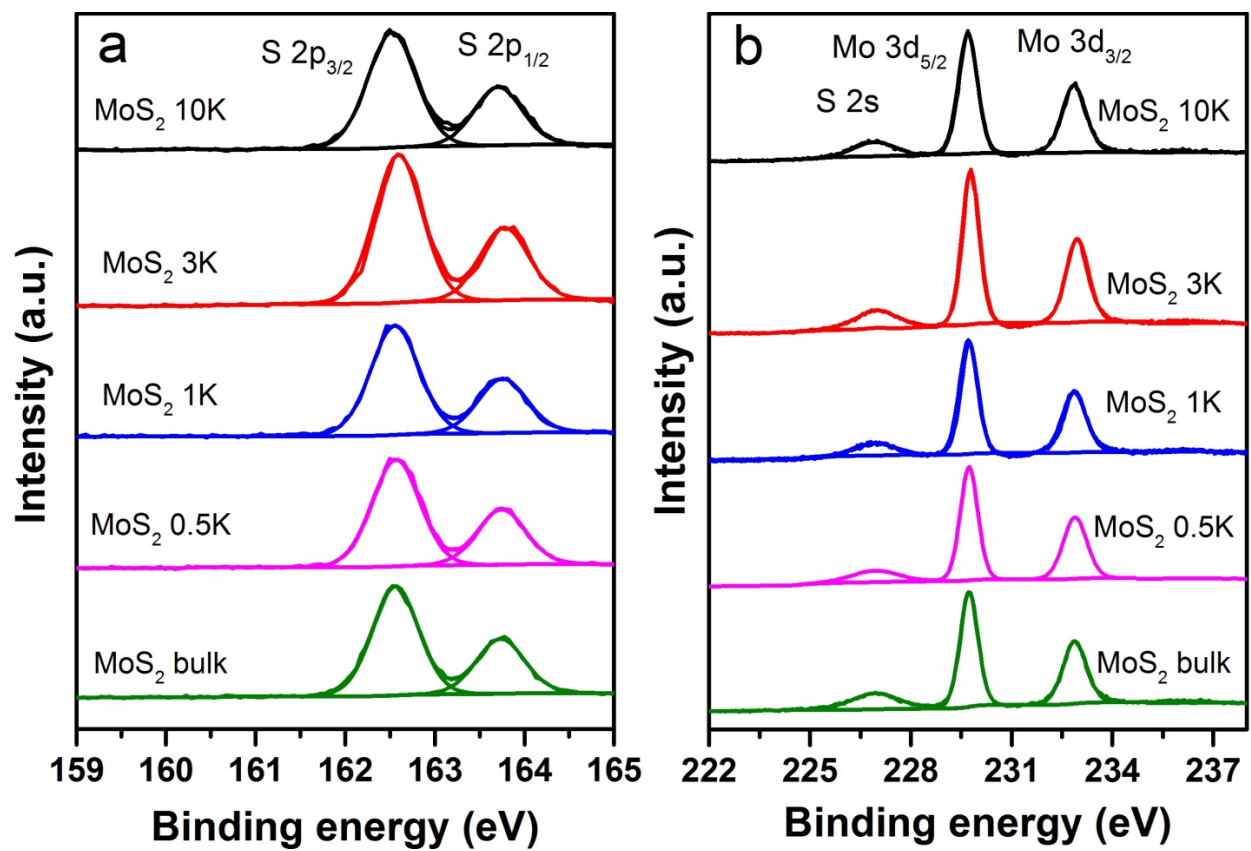

Figure 7. High resolution (a) S 2p and (b) Mo 3d XPS spectra of the five catalysts. S and Mo contents were calculated to have a ratio of 2:1, corresponding to $MoS_2$ phase.

The elemental composition of the five catalysts was characterized using X-ray photoelectron spectroscopy (XPS). The wide survey scan of bulk and $MoS_2$ 10K can be seen in Figure S2a (supporting information). The clear presence of C and O elements for all catalysts is associated with adventitious impurities, which originate from the solvent[53] and the atmosphere. Stoichiometric ratios of S to Mo calculated from the respective integrated areas are close to 2:1 (1.95±0.05:1) demonstrating the expected $MoS_2$ phase. As shown in Figure 7a-b, all the sediments exhibited almost the same binding energies for well-defined S and Mo doublets as those of $MoS_2$ crystal. The peaks around 162.0eV and 163.2eV correspond to S $2p_{3/2}$ and S $2p_{1/2}$ orbitals respectively (Figure 7a), while the peaks at 229.2eV and 232.4 eV are attributed to Mo $3d_{5/2}$ and Mo $3d_{3/2}$ orbitals respectively. These peak positions are indicative of $Mo^{4+}$ and $S^{2-}$ oxidation states in 2H phase of $MoS_2$[12,25]. Binding energies of S 2p and Mo 3d regions remained constant before and after grinding, indicative of no changes in the oxidation states.

The FWHM of Mo $3d_{5/2}$ peak increased from 0.60 to 0.67 for $MoS_2$ bulk and $MoS_2$ 10K respectively. Similarly the FWHM of S $2p_{3/2}$ was increased from 0.59 for bulk to 0.67 at a centrifugation speed of 10K rpm. The atomic concentration of O 1s increased only by 2.31 at% from bulk (9.60 + 0.59 at. %) to $MoS_2$ 10K (11.91 + 0.84 at. %) (Figure S2b and Table S2). This data indicates only a minute transformation of sulfide to oxide and Mo to Mo-O with progressive isolation of smaller and thinner flakes. Furthermore, no noticeable peaks from either Mo higher oxidation states ($Mo^{5+}$ or $Mo^{6+}$) at higher energies of ~236 eV, nor from S higher oxidation states, in the energy region 168 -170 eV, were observed[25], confirming that there is no obvious oxidation of $MoS_2$ nanodots. This is different from the majority of studies on $MoS_2$ nanodots or nanosized flakes, where a significant change in the oxidation state occurs upon exfoliation[33,53,54]. The possible reason is that the ionic liquid protects the sheets and inhibits oxidation of $MoS_2$ during exfoliation. This is similar to our previous work[30], where we reported the production of low oxygen content, a few layer graphene nanosheets by RTIL assisted grinding exfoliation. The presence of predominant low oxidation states of Mo might play a role on the electron filling of bonding and antibonding between the active sites and H, lowering the H bonding energy and activation barrier and thus improving the HER activity.

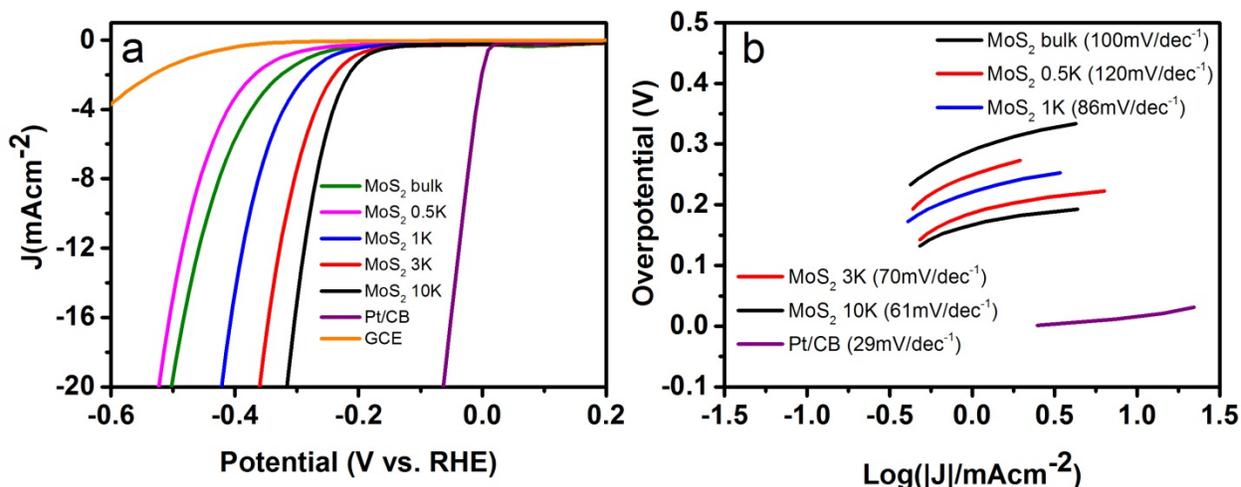

Figure 8. a) Polarisation cures of all five catalysts. GCE and Pt/CB (20 wt. % Pt) are used as comparison. b) Tafel slopes of all five catalysts.

The HER activities of the five $MoS_2$ catalysts were investigated in 0.5 M $H_2SO_4$ in a three-electrode setup, using a GCE electrode (3 mm diameter) with a mass loading of 0.28 mg/cm$^2$. Comparison of polarization curves is provided in Figure 8a together with those of glassy carbon electrode and Pt/C catalyst (20 wt. % Pt). An iR correction has been employed to compensate for any potential loss arising from external resistance of the electrochemical system in this figure as explained in the supporting information (electrochemical measurements). It can be clearly seen that the catalytic activities of $H_2$ generation increased with the decrease in lateral size and number of layers, upon sequential centrifugation. The $MoS_2$ 10K in particular, shows the best catalytic performance exhibiting an onset potential of -0.09 V vs. RHE and an overpotential ($\eta_{10}$) of -248 mV at the cathodic current density of 10 mA/cm$^2$, the last is commonly used as a figure of Merit for comparing the HER performance of electrocatalysts[14,55]. Both potentials are considerably more positive than those of $MoS_2$ bulk displaying an onset potential of -0.18 V and $n_{10}$ = -471 mV. $MoS_2$ 0.5K shows inferior characteristics compared to the bulk due to the absence of fine and small flakes in agreement with the SEM, XRD, and Raman results.

It is well understood that the electronic structure of catalysts can affect HER catalytic activity, as it will affect the Gibbs free energy for hydrogen adsorption on the catalyst and the electron transfer between the catalysts, the reagents and the electrode. By reducing the dimensions both along the in-plane and vertical directions a considerable number of metallic edges becomes available for hydrogen adsorption and at the same time an easier path for transport of electrons is availed, all of which contribute to the improved catalytic activity for HER. The increase in the number of active sites for $MoS_2$ 10K nanodots compared to other catalysts was confirmed by the observed larger double layer capacitance ($C_{dl}$), i.e. the effective electrochemically active surface area[56]. Cyclic voltammograms (CVs) were collected in the region of 0.33–0.13 V, where the current response should only be due to the charging of the double layer (Figure S3 & S4). The capacitance of $MoS_2$ 10K is 3.5 times larger than that of $MoS_2$ bulk.

Tafel slope is a useful metric to assess the performance of catalysts and at the same time is a valuable indicator of the mechanistic reaction processes of HER[57-59]. The linear portions of Tafel plots were fitted to the Tafel equation ($\eta = b log|J| + a$, where $\eta$ is overpotential, $J$ is the current density, $a$ is exchange current density and $b$ is the Tafel slope) and yielded the Tafel slope (Figure 8b). A small Tafel slope is preferred as it indicates a fast increase of hydrogen generation rate with the applied overpotential. Tafel slopes for all catalysts are summarized in Table S3 and are plotted in Figure 8b, which clearly show a trend from high to low slope as the lateral size and thickness decreases. The Tafel slope of 61 mVdec$^{-1}$ for $MoS_2$ 10K is superior to several published works, listed in Table 1, such as $MoS_2$ nanodots prepared by sonication (Ji et al[60], onset = -0.12 V, Tafel = ~70 mVdec$^{-1}$), mechanical (Varrla, et al[61], Tafel = 115 mVdec$^{-1}$) or electrochemical methods (Gopalakrishnan et al[34], onset = ~-0.21 V (vs. SHE), Tafel = 60 mVdec$^{-1}$). To the best of our knowledge, $MoS_2$ nanodots synthesized here are the most efficient $MoS_2$ with 2H phase. The low oxidation state of the $MoS_2$ 10K nanodots achieved by our synthesis process is one important factor contributing to their superior performance. It is well established that edge oxidized 2H phase nanosheets show a dramatically suppressed performance (slope 186 mVdec$^{-1}$) caused by the deactivation of edges[33,53,54].

**Table 1**. Comparison of HER activity of $MoS_2$ 10K with other relevant reported findings.

| HER catalyst | Onset potential (V) | Tafel slope (mVdec$^{-1}$) | Reference |
|---|---|---|---|
| **$MoS_2$ 10K** | -0.09 (0.19 at 1 mAcm$^{-2}$) | 61 | This work |
| **$MoS_2$ composite (NMP)** | -0.12 | 69 | [33] |
| **$MoS_2$ nanosheets** | -0.27 (at 1 mAcm$^{-2}$) | 115 | [61] |
| **$MoS_2$ dots on Au** | -0.16 | 82 | [24] |
| **$MoS_2$ embedded in ordered mesoporous carbon** | (~)-0.12 | 60-65 | [62] |
| **$MoS_2$ dots/nanosheet hybrid on Au** | -0.19 | 74 | [31] |
| **$MoS_2$ nanoparticles on Au** | -0.09 | 69 | [25] |
| **Li-$MoS_2$** | -- | 62 | [63] |
| **$MoS_2$ composite (NMP)** | -0.12 | 69 | [33] |
| **$MoS_2$ nanoplate assemblies** | -0.09 | 68 | [48] |
| **NSs-550** | -- | 68 | [64] |
| **$V_{0.09}Mo_{0.91}S_2$ (Vanadium doped $MoS_2$)** | -0.13 | 69 | [65] |

According to the classic theory, hydrogen evolution, proceeds through three principle reaction steps in acidic media[57,66]:

$$H_3O^+ + e^- \rightarrow H_{ad} + H_2O \quad \text{Volmer reaction (Tafel slope 120 mVdec}^{-1}\text{)} \quad (1)$$

$$H_{ad} + H_3O^+ + e^- \rightarrow H_2 + H_2O \quad \text{Heyrovsky reaction (Tafel slope 40 mVdec}^{-1}\text{)} \quad (2)$$

$$2H_{ad} \rightarrow H_2 \quad \text{Tafel reaction (Tafel slope 30 mVdec}^{-1}\text{)} \quad (3)$$

Where $H_{ad}$ represents the hydrogen adsorption sites onto the surface of a metal catalyst. According to theory, the slopes associated with Volmer, Heyrovsky and Tafel reactions are 120 mVdec$^{-1}$ (equation 1), 40 mVdec$^{-1}$ (equation 2) and 30 mVdec$^{-1}$ (equation 3), respectively. Experimentally two main pathways are generally observed for HER, Volmer−Heyrovsky reaction (equations 1 and 2) and Volmer−Tafel reaction (equations 1 and 3)[67,68]. In the present study, the Tafel slope of 61 mVdec$^{-1}$ for MoS$_2$ 10K suggests that HER was most probably controlled by both electron reduction of protons, which provides a hydrogen atom bound to an active site (equation 1) and electrochemical desorption of hydrogen (equation 2) (Volmer−Heyrovsky reaction). In contrast for crystalline MoS$_2$ bulk, the HER proceeds through to the Volmer reaction suggested by its large Tafel slope of 100 mVdec$^{-1}$. For both MoS$_2$ bulk and MoS$_2$ 0.5K (120 mVdec$^{-1}$) the reaction kinetics are limited by the inefficient number of edges, where adsorption of $H^+$ takes place, as indicated by the large Tafel slope[14]. At higher centrifugation speeds, the lower Tafel slopes suggest that the number of accessible active sites on the 2H phase MoS$_2$ nanosheets have increased.

The intrinsic activity of the catalyst materials was studied using the turnover frequency (TOF), which represents the number of hydrogen molecules produced per second per active site[23,69].

TOF of $MoS_2$ 10K at -0.2 V (3.3 $H_2$/s per active site) was more than four times higher than that of $MoS_2$ bulk (0.7 $H_2$/s per active site), highlighting the important influence of exfoliation and small lateral size in HER activity (Figure 9). The TOF value obtained for $MoS_2$ 10K shows a turnover frequency (TOF) of 3.3 $H_2$/s at an overpotential of -0.2 V, thus outperforming $MoO_3$-$MoS_2$ nanowires[70] (~0.7 $H_2$/s at -0.2V) and double gyroid $MoS_2$[23] (~0.7 $H_2$/s at -0.2V) as shown in Table S4. These results reveal that the combination of decreased particle size, thickness and high crystallinity is highly beneficial for the greatly enhanced electrocatalytic activity and the stability of $MoS_2$.

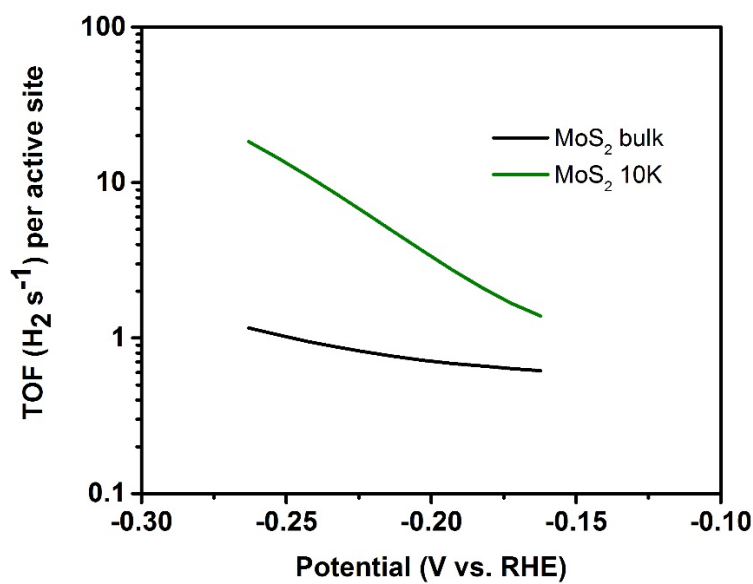

Figure 9. Turnover frequencies of $MoS_2$ 10K and $MoS_2$ bulk. TOF calculations were conducted per Mo atom.

One should note that a critical issue in calculating accurately TOF originates from the precision in measuring the number of active sites (n) and the electrochemically active surface area (ECSA). The above TOF calculations were conducted assuming that the surface exposed Mo atoms are the active sites. However, for crystalline 2H-$MoS_2$, only exposed edge sites contribute to HER activity. Such an uncertainty would lead to inaccuracies in the magnitude of the actual TOF.

The estimation of electrochemically active surface area was conducted by measurement of the double-layer capacitance in a potential region with no faradaic response following McCory's et al. methodology[71]. The ESCA was estimated from the ratio of the measured double layer capacitance with respect to the specific capacitance of an atomically smooth $MoS_2$ material (~60 $\mu F/cm^2$).

$$ECSA = \frac{C_{dl}(mFcm^{-2})}{C_{dl}(\mu Fcm^{-2})} \qquad (4)$$

Using equation (4), we have obtained values of 35.83 and 10.32 for the ESCA of $MoS_2$ 10K and $MoS_2$ bulk respectively (Table S3). However, this method although suitable for electrodes consisting of conductive materials, could lead to error in the ECSA determination for semiconducting layers, such as $2H-MoS_2$, where an increase in active surface area does not necessarily translate into an increase in the double layer capacitance. These uncertainties in the estimation of both n and ESCA in equation (S2) can lead to estimated TOFs that differ by orders of magnitude. These factors make it difficult to conduct meaningful comparisons of TOFs with the literature. In addition, for different MoSx-based catalysts, the comparison of the TOFs is only meaningful when the value is taken at the same overpotential.

.

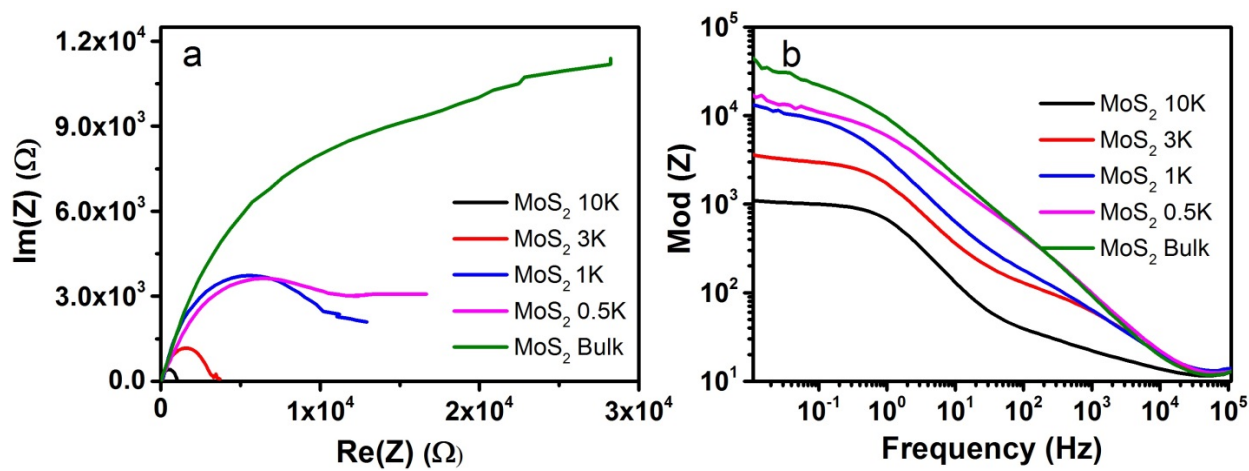

Figure 10. EIS plots in the form of (a) Nyquist and (b) Bode plots of the five catalysts Measurements were performed at -0.22 V vs. RHE.

The electrode kinetics under catalytic HER operating conditions (Figure 10) were investigated by electrochemical impedance spectroscopy (EIS). The impedance spectra mirror the HER activity, by applying a small voltage (-0.22 V) close to the onset potential. A high HER activity is reflected by a small semicircle in the Nyquist plot (Re(z) vs Im(z)) over the frequency range 1 MHz to 10 mHz, which indicates a small charge transfer resistance (Rct) as it allows fast shuttling of electrons during HER (Figure 10a). This is obvious in Bode plot, which presents the modulus of impedance $|Z| = \sqrt{|Re(Z)|^2 + |Im(Z)|^2}$ as a function of log of frequency (Figure 10b). In the low frequency regime it is obvious that modulus of Z is the smallest for MoS$_2$ 10K nanodots. It is clear that as exfoliation products are reduced both in lateral size and thickness the Rct dramatically decreases. This trend is consistent with those of Tafel slopes and polarization curves. These electrochemical data, together with the structural characterization results, consistently show a dramatic increase in intrinsic electrocatalytic activity for HER as the products evolve from large and thick platelets to a few layer nanodots.

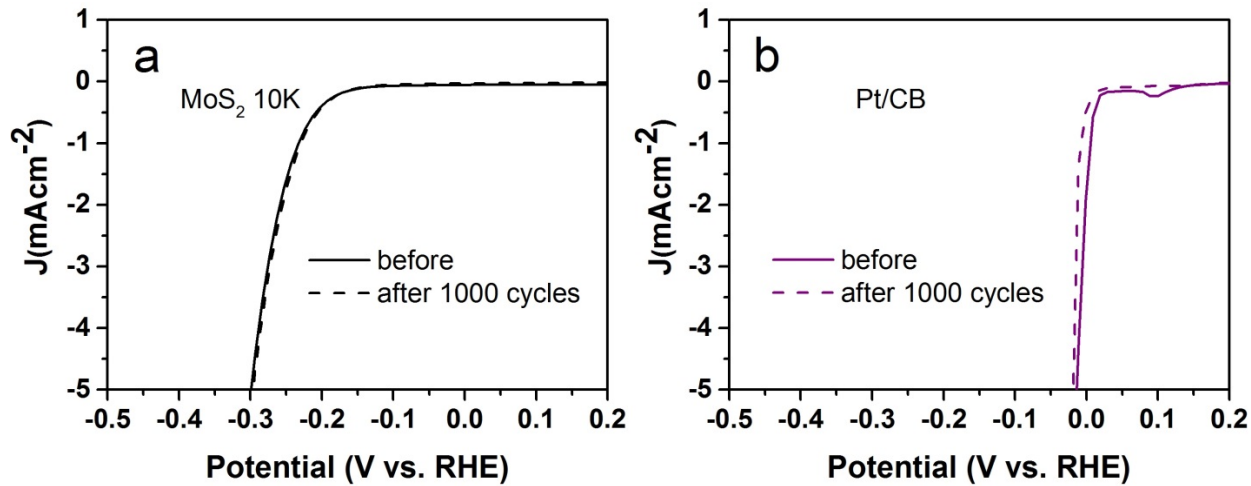

Figure 11. Polarization curves of a) $MoS_2$ 10K and b) commercial Pt/CB (20 wt. % Pt) before and after 1000 CV scans. All measurements were carried out in 0.5M $H_2SO_4$ using a 3 mm diameter BASI GCE and a catalyst loading of 0.283 mg/cm$^2$.

The activity of both $MoS_2$ 10K and commercial Pt/CB (20 wt % Pt) before and after application of 1000 cycles (between +0.2 to -0.4 V vs RHE at a scan rate of 100 mVs$^{-1}$) is presented in Figure 11a-b. No measurable change was observed for $MoS_2$ 10K. In contrast the Pt/CB exhibited a noticeable shift after 1000 cycles. The results reveal the exceptional stability of $MoS_2$ 10K, which are in agreement with the well reported stability of crystalline 2H-$MoS_2$. This is linked to the ability of crystalline $MoS_2$ to resist oxidation in contrast to amorphous $MoS_2$, which is more susceptible to oxidation[72]. However, it should be noted that intense stability tests in excess of 10,000 potential cycles are required for real working devices.

**Conclusions**

In summary, we demonstrate the exfoliation of $MoS_2$ sheets via a simple ionic liquid assisted grinding method combined with sequential centrifugation steps. $MoS_2$ nanodots having a thickness of up to 7 layers (~4 nm) and an average lateral size smaller than 20 nm were isolated through sequential centrifugation steps. A drastic increase in catalytic $H_2$ evolution was observed with a simultaneous decrease in number of layers and later size. The obtained $MoS_2$ nanodots exhibit an onset potential as small as -0.09 mV, a high exchange current density of $1.6 \times 10^{-3}$ $mAcm^{-2}$, and a Tafel slope of 61 $mVdec^{-1}$. Moreover, they merely require overpotentials as low as -248 mV to attain current densities of 10 $mAcm^{-2}$ and show good long-term stability. All these characteristics comparable favorably to all recently reported $MoS_2$ quantum dots or nanosheet catalysts having 2H phase. The improved catalytic activity can be attributed to the following factors (i) the high abundance of edge sites exposing S edge atoms, which play a crucial role in catalyzing the electrochemical HER and (ii) the high degree of exfoliation (reduced number of layers,) which allows efficient electron transfer between the electron rich edges and the electrode (iii) the low oxidation state of the $2H-MoS_2$ nanodots, which substantially improves the charge transfer kinetics of HER. We have demonstrated that the HER performance of $MoS_2$ varies significantly with size and thickness reduction and hence provides fundamental insight for future theoretical and experimental research. The novel fabrication method reported here is scalable and can be extended to obtain other 2D layered based catalysts and therefore represents an important development toward water splitting and other energy conversion technologies.

**Corresponding Author**

*To whom correspondence should be addressed, p.papakonstantinou@ulster.ac.uk

**Acknowledgements:** This work was financially supported by a PhD Studentship to J.B from the Department of Employment and Learning in Northern Ireland; A visiting Senior Research Fellowship to M.L. by Ulster University; The Chinese thousand-talents plan program and the Jiangsu Shuangchuang program (P. W.).

# Supporting Information

# Electrocatalytic Hydrogen Evolution reaction on edges of a few layer Molybdenum disulfide nanodots

*John Benson[†], Meixian Li[±], Shuangbao Wang[§], Peng Wang[§] and Pagona Papakonstantinou[†]\**

[†]School of Engineering, Engineering Research Institute, Ulster University, Newtownabbey BT37 0QB, UK

[±]College of Chemistry and Molecular Engineering, Peking University, Beijing 100871, P.R.China.

[§]National Laboratory of Solid State Microstructures, College of Engineering and Applied Sciences and Collaborative Innovation Center of Advanced Microstructures, Nanjing University, 22 Hankou Road, Gulou, Nanjing, 210093, P. R. China

*To whom correspondence should be addressed, p.papakonstantinou@ulster.ac.uk

## I. Electrochemical Measurements

Catalyst inks were prepared by dispersing 5 mg of catalyst material in 1 ml DMF and ultrasonicated for 1 hour. 50 μl of Nafion solution was then added and sonicated for a further 10 minutes. 4 μl of catalyst ink was deposited onto a polished glassy carbon electrode (GCE, 3 mm, BASI) for a catalyst loading of 0.283 mg/cm² and dried under an IR lamp. Electrochemical measurements were performed in 200 mL of 0.5 M $H_2SO_{4(aq)}$ solution employing a three-electrode configuration and an Autolab, PGSTAT20/FRA system. A platinum wire was used as a counter electrode and an Ag/AgCl (3M KCL) electrode was used as a reference electrode. Conversion to RHE was calculated using equation (S1)[1]. Polarisation curves were performed under ambient conditions under potentials between -0.8 and +0.2 V vs Ag/AgCl at a 10 mVs$^{-1}$ scan rate. Working electrodes were pre-conditioned prior to polarisation curves by performing cyclic voltammetry at a scan rate of 100 mVs$^{-1}$ for 20 scans.

$$E_{RHE} = E_{Ag/AgCl} + 0.0591 * pH + 0.1976 \tag{S1}$$

Electrochemical impedance spectra (EIS) were measured with the working electrode biased at -0.22 V (vs RHE) and superimposing a small sinusoidal voltage of 10 mV over the frequency range 1 MHz to 10 mHz.

Polarization curves from all catalysts were iR corrected, where the R is the ohmic resistance arising from the external resistance of the electrochemical and was measured by Electrochemical impedance spectroscopy (EIS). As R was taken the impedance value at the high frequencies (10$^5$ Hz) of the bode plot (Figure 10b).

To evaluate the double-layer capacitance ($C_{dl}$) cyclic voltammograms were acquired in a non-faradaic region between 0.13 V to 0.33 V (vs. RHE) at various scan rates (0.01, 0.02, 0.04, 0.06, 0.08 and 0.1Vs$^{-1}$ vs RHE). The C$_{dl}$ was calculated from the slope of the straight line (($i_c = vC_{dl}$)), when the charge current density (J(mAcm$^{-2}$)) at a particular potential is plotted against scan rate.

TOF calculations were carried out using the equations outlined in equation S2$^2$.

$$TOF = \frac{JN_A}{2FnECSA} \qquad (S2)$$

where J is the current density, 2 represents the stoichiometric number of electrons consumed in the electrode HER reaction, ESCA is the electrochemically active surface area of the electrode, F is the Faraday constant (F= 96485 C mol$^{-1}$), N$_A$ is the Avogadro's number (N$_A$=6.022 10$^{23}$ H$_2$ molecules/mole) and n is the number of active sites (n= 1.28 × 10$^{14}$/cm$^2$)$^3$ in a flat 1 cm$^2$ surface of MoS$_2$ sample. Figure 9 shows the TOF values of both bulk and MoS$_2$ 10K in the applied potential region of 0.2 to 0.4V vs RHE, where HER is kinetically controlled. We note that TOF calculations were conducted per Mo atom (as opposed to S atom) to facilitate comparison of different catalysts[4].

II.  **Characterisation Methods**

The low-magnification TEM images of as-prepared catalysts were taken on a FEI TECNAI TF20 TEM at an accelerating voltage of 200 KV. The high resolution lattice fringe images were also taken on TF20 TEM but with relative weak low-dose electron beam to minimize radiation damage. TEM samples were prepared by dropping 2 μl of well dispersed catalyst in DMF onto carbon micro-grids (Agar scientific, S147-3, holey carbon film 300 mesh Cu). The grids were then dipped into acetone to remove excess DMF and dried under ambient conditions. High-resolution XPS spectra were taken using a Kratos AXIS ultra DLD with an Al Kα (hv=1486.6 eV) x-ray source. Spectra were fitted to a Shirley background. Raman spectra were obtained with a Labram 300 system using a He-Ne (632.8nm) laser. XRD was conducted on powdered samples with a Bruker AXS D8 discover with Cu-kα radiation (40kV, 20mA, λ = 1.5418Å). SEM images were taken on a Quanta 200 3D system. UV-vis spectra were taken on a Perkin Elmer Lamba 35 spectrometer. Spectra were taken in the range of 200 nm to 800 nm in DMF solution.

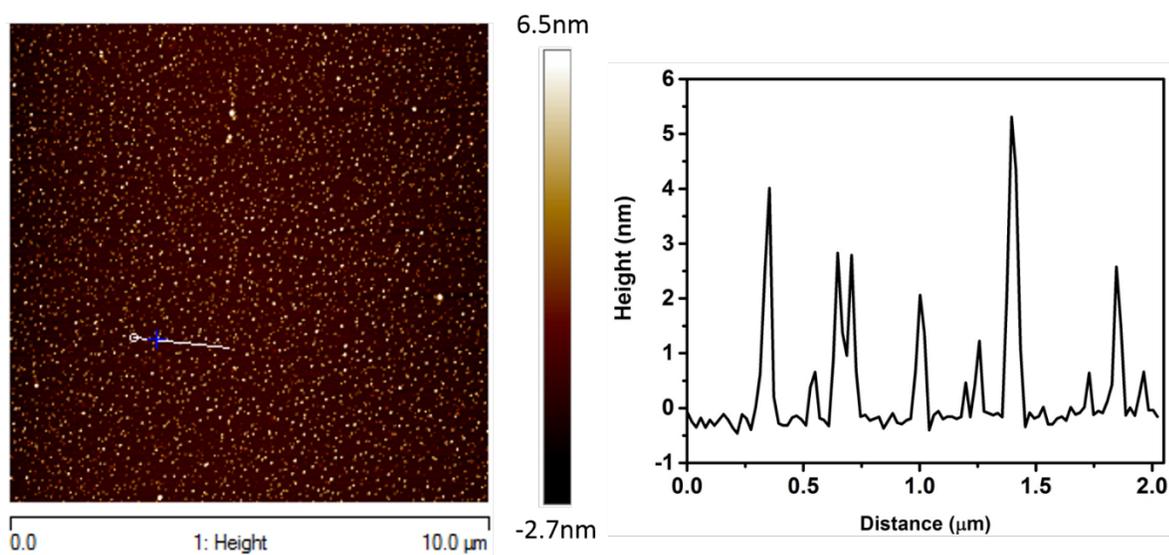

Figure S1. AFM image showing thickness up to 7 layers are present in MoS 10K.

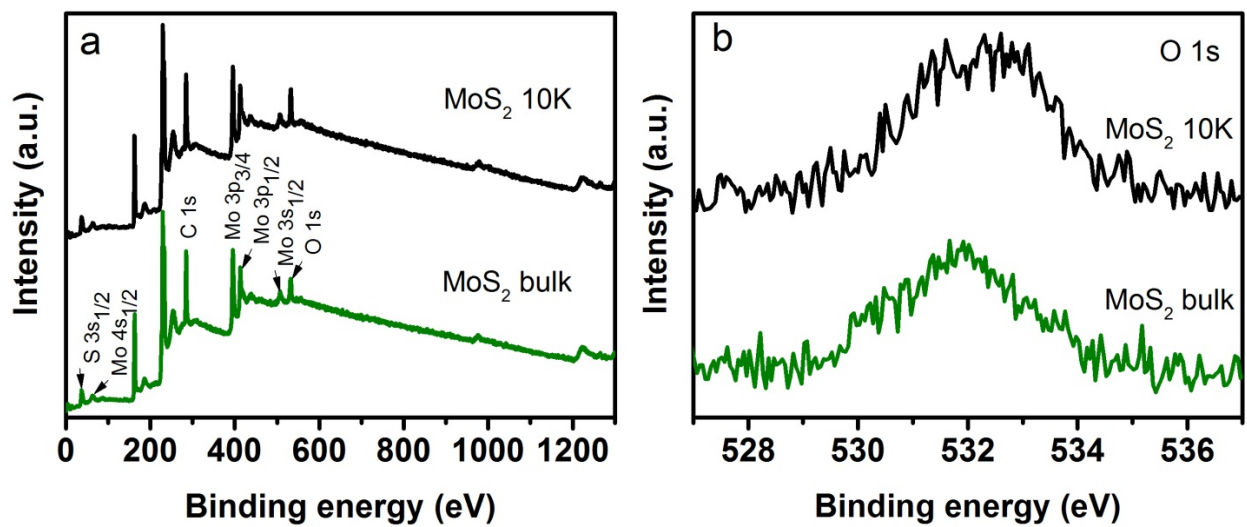

Figure S2. XPS (a) wide survey and (b) O 1s high resolution spectra of $MoS_2$ bulk and $MoS_2$ 10K.

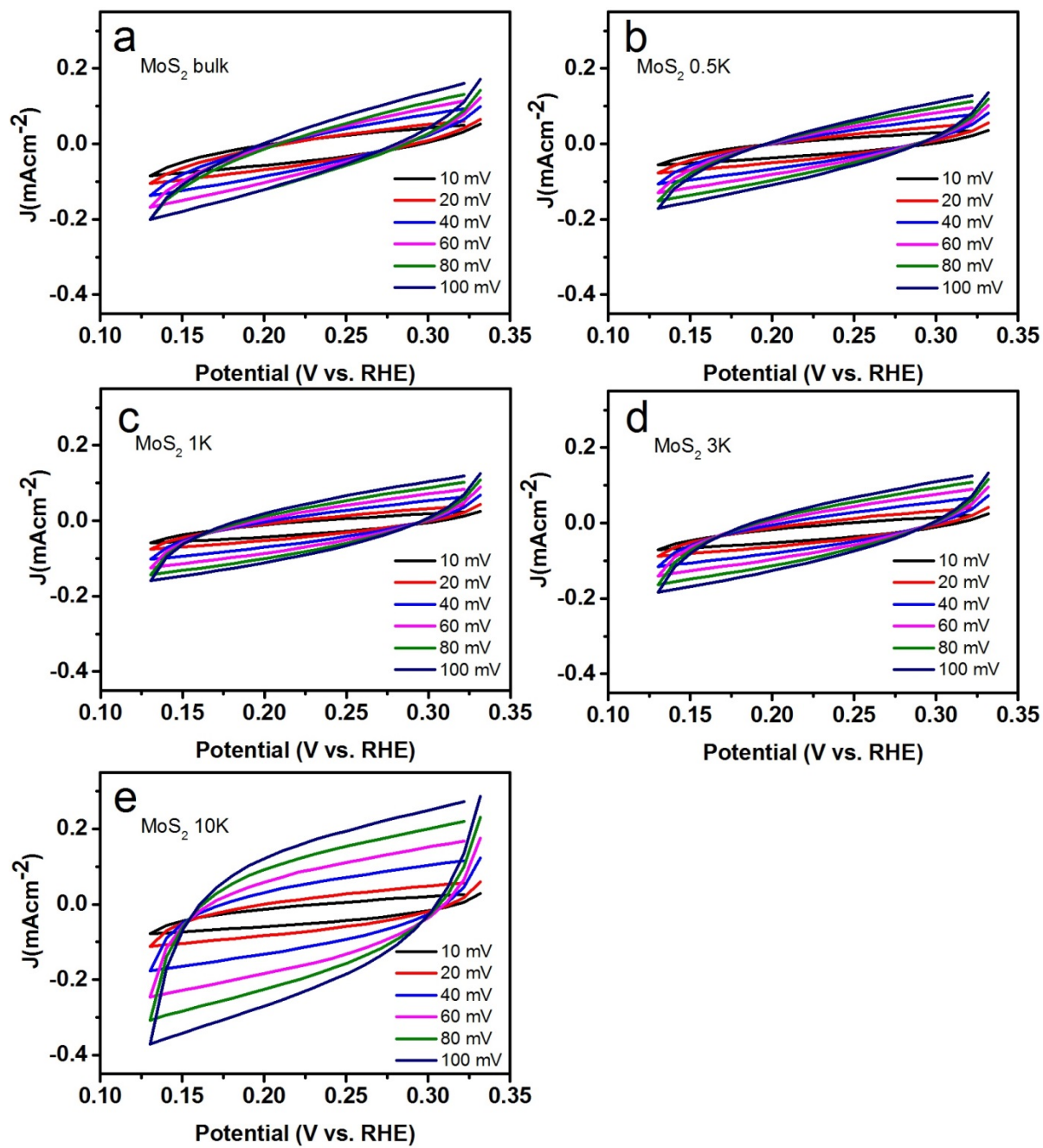

Figure S3. Cyclic voltammograms of (a) MoS$_2$ Bulk, (b) MoS$_2$ 0.5K, (c) MoS$_2$ 1K, (d) MoS$_2$ 3K and (e) MoS$_2$ 10K. Used to measure a non-faradic region between +0.13 to +0.33 V (V vs. RHE) at various scan rates (0.1, 0.08, 0.06, 0.04, 0.02 and 0.01 Vs$^{-1}$).

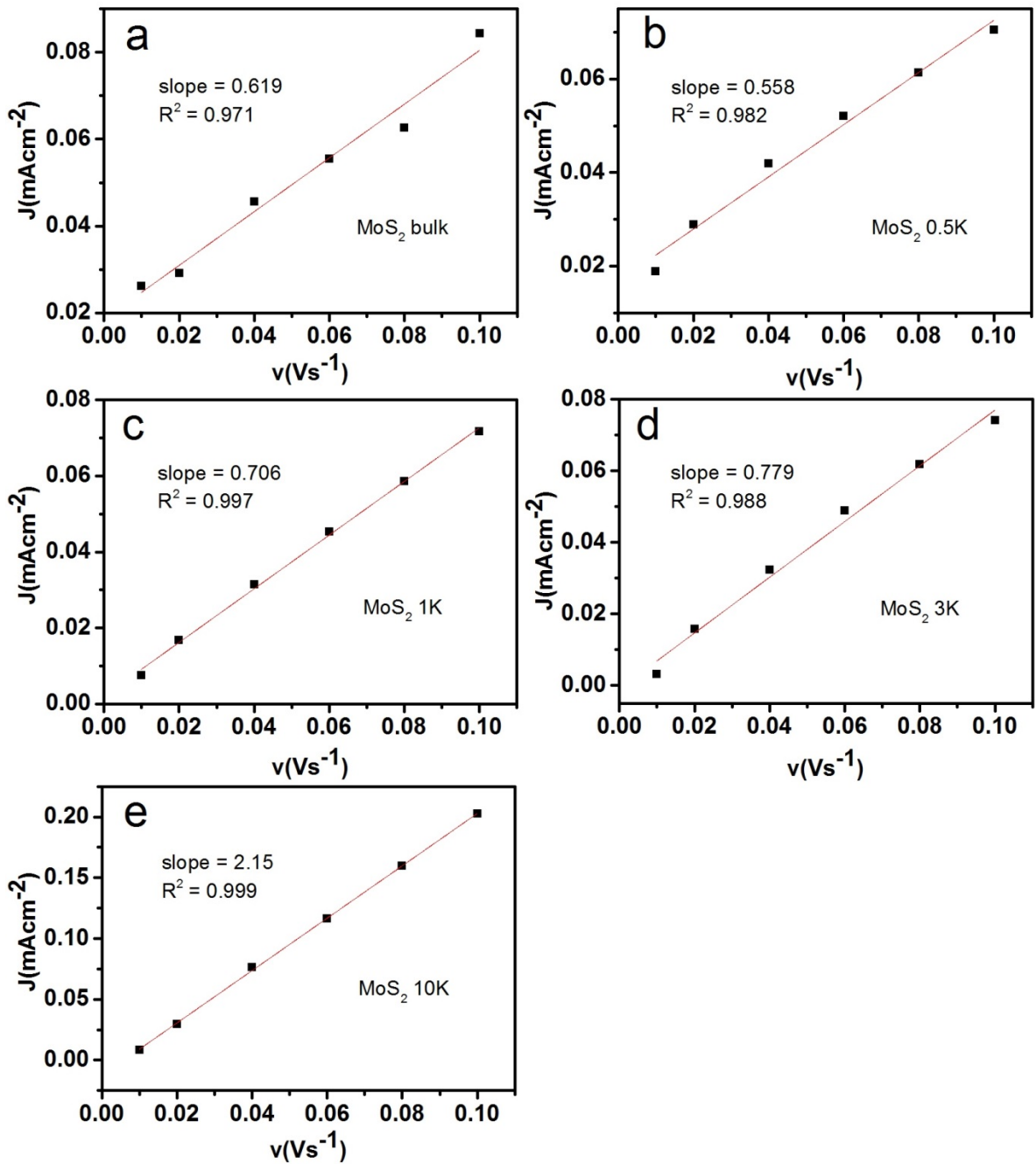

Figure S4. Capacitive current density measured at 0.025 V plotted as a function of scan rate for (a) bulk $MoS_2$, (b) $MoS_2$ 0.5K, (c) $MoS_2$ 1K, (d) $MoS_2$ 3K and (e) $MoS_2$ 10K. The average value of the slope was determined as the double-layer capacitance ($C_{dl}$) of each catalyst.

Table S1. Summary of FWHM values of the (002) plane from all five catalysts, Measurements conducted by XRD.

|  | $MoS_2$ bulk | $MoS_2$ 0.5K | $MoS_2$ 1K | $MoS_2$ 3K | $MoS_2$ 10K |
|---|---|---|---|---|---|
| **FWHM (002)** | 0.143° | 0.143° | 0.203° | 0.236° | 0.396° |

Table S2. Elemental composition of all five catalysts.

|  | $MoS_2$ bulk | $MoS_2$ 10K |
|---|---|---|
| **O 1s** | 9.60±0.59 | 11.91±0.85 |
| **C 1s** | 53.57±0.57 | 52.56±0.88 |
| **Mo 3d** | 12.27±0.40 | 11.85±0.13 |
| **S 2p** | 24.56±0.78 | 23.68±0.29 |

Table S3. Electrochemical activity of $MoS_2$ catalysts.

|  | Onset potential (V vs. RHE) | Tafel Slope (mVdec$^{-1}$) | $J_0$ ($\mu A/cm^2$) | $C_{dl}$ (mF/cm$^2$) | ECSA (cm$^2$) |
|---|---|---|---|---|---|
| **MoS bulk** | -0.18 | 100 | $8.7 \times 10^{-4}$ | 0.619 | 10.32 |
| **MoS 0.5K** | -0.18 | 120 | $2.9 \times 10^{-3}$ | 0.558 | 9.3 |
| **MoS 1K** | -0.15 | 86 | $1.7 \times 10^{-3}$ | 0.706 | 11.77 |
| **MoS 3K** | -0.11 | 70 | $5.8 \times 10^{-4}$ | 0.779 | 12.98 |
| **MoS 10K** | -0.09 | 61 | $1.6 \times 10^{-3}$ | 2.15 | 35.83 |

Table S4. Comparison of TOF values.

| Catalyst | TOF (H$_2$/s) at -0.2V | Reference |
|---|---|---|
| **MoS$_2$ 10K** | 3.3 | This paper |
| **MoS$_x$/PP-CFP** | 0.32 | [5] |
| **MoS$_3$-CV** | ~14 at -0.3 V (MoS 10K 39.5 at -0.3V) | [6] |
| **MoS$_x$-G** | ~1.5 | [7] |
| **MoS$_x$-NCNT** | 3.5 | [8] |
| **Wet chemical synthesized MoS$_2$** | 0.3 (based on Mo sites) | [9] |
| **Double gyroid MoS$_2$** | ~0.7 (based on Mo sites) | [10] |
| **[MoS$_3$S$_{13}$]$^{2-}$|GP** | ~0.7 (based on Mo sites) | [3] |
| **MoO$_3$-MoS$_2$ nanowires** | ~0.15 (based on Mo sites) | [11] |